\documentclass[twocolumn]{aastex62}

\newcommand{\ott}{{O$_{32}$}}

\newcommand{\km}{{\rm\thinspace km}}

\newcommand{\s}{{\rm\thinspace s}}

\newcommand{\kmps}{\hbox{$\km\s^{-1}\,$}}

\newcommand{\fesc}{$f_{\rm esc}$}

\newcommand{\ha}{H$\alpha$}

\newcommand{\hb}{H$\beta$}
\newcommand{\hg}{H$\gamma$}
\newcommand{\hd}{H$\delta$}

\newcommand{\hi}{H\thinspace{\sc i}}
\newcommand{\hii}{H\thinspace{\sc ii}}

\newcommand{\oiii}{[O\thinspace{\sc iii}]}

\newcommand{\oii}{[O\thinspace{\sc ii}]}

\newcommand{\ariv}{[Ar\thinspace{\sc iv}]}

\newcommand{\sii}{[S\thinspace{\sc ii}]}

\newcommand{\Mdot}{\hbox{$\dot M$}}

\newcommand{\hei}{He\thinspace{\sc i}}

\newcommand{\solarmass}{M$_{\odot}$}
\usepackage{ulem}

\graphicspath{{./}{figures/}}

\shorttitle{Blow-Away in Pox~186}

\begin{document}

\title{Blow-Away in the Extreme Low-Mass Starburst Galaxy Pox~186}

\email{eggen091@umn.edu}

\author{Nathan R. Eggen}
\affiliation{Minnesota Institute for Astrophysics, University of Minnesota}

\author{Claudia Scarlata}
\affiliation{Minnesota Institute for Astrophysics, University of Minnesota}

\author{Evan Skillman}
\affiliation{Minnesota Institute for Astrophysics, University of Minnesota}

\author{Anne Jaskot}
\affiliation{Department of Astronomy, Williams College}



\begin{abstract}

Pox 186 is an exceptionally small dwarf starburst galaxy hosting a stellar mass of $\sim10^5$ M$_{\odot}$. Undetected in \hi\ (M $ < 10^6$ M$_{\odot}$) from deep 21 cm observations and with an \oiii/\oii\ (5007/3727) ratio of 18.3 $\pm$ 0.11, Pox~186 is a promising candidate Lyman continuum emitter. It may be a possible analog of low-mass reionization-era galaxies. We present a spatially resolved kinematic study of Pox 186 and identify two distinct ionized gas components: a broad one with $\sigma > 400$ \kmps, and a narrow one with $\sigma < 30$ \kmps. We find strikingly different morphologies between the two components and direct evidence of outflows as seen in the high velocity gas. Possible physical mechanisms driving the creation of high velocity gas seen in \oiii\ are discussed, from outflow geometry to turbulent mixing between a hot (10$^6$ K) star-cluster wind and cooler (10$^4$ K) gas clouds. We find a modest mass-outflow rate of 0.022 \solarmass\ yr$^{-1}$ with a small mass loading factor of 0.5, consistent with other low mass galaxies. Finally we compare the mass-loading factor of Pox~186 with extrapolations from numerical simulations and discuss possible reasons for the apparent discrepancy between them.


\end{abstract}



\section{Introduction} \label{sec:intro}

\subsection{On Outlfows}

Star-formation and stellar feedback are primary ingredients in galaxy evolution, with photoionization and momentum deposition into the surrounding interstellar medium (ISM) being the main sources of stellar feedback. In a starburst large enough, the mechanical feedback can be sufficient to drive major outflows, called galactic winds, moving material into the circumgalactic medium (CGM) and intergalactic medium (IGM) beyond.

Models of galactic winds have evolved overtime. Early models predicted that winds in dwarf galaxies have a larger impact as compared to high mass systems \citep{1986ApJ...303..39,1999ApJ...513..142,2000ApJ...313..291}. This is due to the shallower potential wells of dwarf galaxies allowing for more easily launched winds. Since then gas removal, re-accretion of ejected material, inhibition of accreting material due to heating, mixing between gas phases in the outflow, and heating of the ISM have all been added to hydrodynamical simulations \citep{2009MNRAS...395..1875,2016ApJ...824..57,2017ApJ...846..66,2020arXiv...2002.10468}. These studies have provided quantitative predictions on how winds impact galaxy properties such as stellar mass growth, and metallicity in the CGM and IGM \citep{2018ApJ...867..142}. 

Verifying predictions with observations, however, is challenging due to the physical nature of winds. Galactic winds have low surface brightnesses and are transient in nature resulting from being driven by the stellar winds and resulting supernovae of the largest stars. Compounding this effect, outflows are multi-phase phenomena, requiring observations on gas temperatures ranging from 10$^8$ K to molecular gas at $\sim$10$^2$ K to have a full picture. As a result seemingly simple properties of outflows, for example their radial extent, have still not been observationally measured.  

Each phase of the outflow has its own characteristics. The hot phase of the outflow (T$>$10$^6$ K) typically contains $\sim$10\% of the mass \citep{2017ApJ...834..25}, but has the highest velocity and likelihood of reaching the IGM (v$>$1500 \kmps). It also tends to be metal enriched, being made up of supernova ejecta and thermalized stellar winds. Most mass tends to be contained in the warm phase (T$\sim$10$^4$ K). The warm phase consists of ionized gas in the ISM swept up by the outflow, cooler shock-heated gas, and condensed gas that cooled out of the hot phase. It tends to have lower velocities than the hot phase ($v\sim$500-1000 \kmps) and, depending on the galaxy, can also reach the CGM and IGM. Finally the cold neutral gas phase (T$<$10$^3$ K) has been found to be ubiquitous in galactic outflows of high mass galaxies, making up as much as 50\% of the outflowing mass in the most extreme cases. In models of dwarf galaxies, however, the cool phase mass is subdominant to the warm phase \citep{2012MNRAS...421..3522,2015MNRAS...454..2092,2016ApJ...824..57,2019MNRAS...483..3363,2019MNRAS...490..3234}. 

The warm phase tends to be the easiest phase to observe due to the plethora of nebular lines emitted by the gas in the optical, but care must be taken to separate the emission in the outflow from the host galaxy. This disentanglement turns out to be a major problem. Simple imaging of the outflowing gas is dubious, where nebular lines from the outflow are dominated by the starburst in the host galaxy. With kinematic information the high velocity gas can be isolated, but long-slit and multi-object spectrographs may provide only partial spatial information. Integral field spectroscopy therefore is a natural solution for the study of outflows, providing both spectral and spatial information simultaneously. Even so, projection effects complicate the interpretation of the data. Despite these challenges, it is possible to learn much about galactic winds. 

A common method of comparison between outflows of different galaxies and simulations is to measure the mass outflow rate (\Mdot) relative to the star formation rate (SFR). This quantity is known as the mass-loading factor ($\eta$ = \Mdot/SFR) and is a measure of the efficiency of the outflow. Observations have found mass-loading factors ranging widely, from less than 1 to greater than 50 in the most extreme cases \citep{1999ApJ...513..156,2015ApJ...809..147,2017MNRAS...469..4831}. Simulations also find widely varying mass-loading factors \citep{2015MNRAS...454..2691,2016ApJ...824..57,2019MNRAS...483..3363,2019MNRAS...490..3234}, but where observations are conflicting if there is a trend between stellar mass and mass-loading factor, simulations agree, finding galaxies with lower stellar masses driving more efficient outflows \citep{2019ApJ...886..74}. 

Variations in mass-loading factors between galaxies are to be expected. The timescale intrinsic to the method of measuring mass-outflow rates gives an instantaneous snapshot of the outflowing material. Star-formation rates, however, measure the amount of star formation over the last few 10s of Myrs, during which time an outflow and the instantaneous SFR can evolve significantly. This difference in timescale can manifest as changes in measured mass-loading factors depending on the specific scenario in a given galaxy. A simple case is the difference between an instantaneous starburst and continuous star-formation. The instantaneous starburst would be expected to produce a more powerful outflow from the focused feedback, but the measured SFR would be the same. Another interesting scenario that would affect the mass-loading factor is an ISM blow-away, where a significant fraction of the ISM is removed from the galaxy through feedback\footnote{We adopt the definitions from \citealt{1999ApJ...513..142} when referring to blow-outs and blow-aways.}. If \Mdot\ is measured a sufficiently long time after the initial burst of star-formation to allow the blow-away to occur, then there would be less gas remaining to be ``loaded” into the outflow, and the mass outflow rate would be suppressed.

\subsection{Outflows, Reionization, and Green Pea Galaxies}

Outflows are also important in the context of escaping ionizing radiation from the host galaxy. The low-density, highly ionized channels created by outflows in the ISM are ideal environments for ionizing radiation to escape, which is a major problem in the study of reionization.

Low mass star-forming galaxies are thought to be the primary sources of ionizing photons during the epoch of
reionization (\citealt{2017AAS...835..113,2018ApJ...869..123}; see \citealt{2020ApJ...892..109} for an alternative model). A galaxy's capacity to contribute to
reionization depends on the total ionizing flux produced and the fraction that escapes its
interstellar medium (\fesc), reaching the intergalactic medium (IGM, \citealt{2015ApJ...802..L19}). For
reionization to complete by z$\sim$6, \textit{all} galaxies must have had an average escape fraction between
10-20\% \citep{2015ApJ...810..72,2015ApJ...802..L19} or the ionizing photon production efficiency must
be higher than previously assumed \citep{2019ApJ...879..36}. 

An increasingly opaque IGM, however, makes direct observations of these galaxies' escaping ionizing flux
difficult. Beyond z$\sim$4, the number of neutral hydrogen cloud absorbers becomes large enough to inhibit
direct observations of ionizing flux. Observations have found the evolution with redshift of the number of absorbers above a certain column density can be described as a broken powerlaw \citep{1980ApJ...42..41,1982ApJ...48..455,1993ApJ...418..28},
with the exponent increasing from 2.82 to 5.07 at $z$ $>$ 3.11. With such a sharp increase in the number of absorbers at
high redshift, determining the amount of ionizing radiation an object emits is increasingly difficult to
reconstruct from the amount of ionizing radiation that reaches an observer. 
Therefore, a set of observables that indicate high escape fractions but are also accessible at z$ > 6$ is
needed to further constrain the escaping ionizing flux from galaxies responsible for reionization. 

Galaxies with high escape fractions have been found among low redshift analogs, notably the so-called ``Green Peas" (GPs,
\citealt{2009MNRAS...399..1191}). Selected for their high \oiii/\oii\ (5007/3727; \ott) ratio, Green Peas are compact, highly
star-forming galaxies similar to those responsible for reionization. Some Green Peas
have been found with escape fractions as high as 70\% \citep{2016aNature...529..178,2016bMNRAS...461..3683,2016A&A...591..L8,2017A&A...597..A13} and a strong anti-correlation between the separation of Lyman alpha peaks and \fesc\
has been observed \citep{2018MNRAS...478..4851}, implying gas kinematics and structure play an important role in
determining \fesc. Lyman alpha's resonance, however, makes its line-shape difficult to interpret. In
addition, Green Peas are limited as useful analogs if indeed low-mass galaxies dominate reionization. They are many times more massive than these galaxies
(e.g., \citealt{2017MNRAS...469..L83}) while simultaneously being too compact and
far away (z$\sim$0.3) to be structurally studied in detail. Looking locally (z$\sim$0) allows for the mapping
of gas kinematics in galaxies with properties analogous to Green Peas, and masses similar to galaxies thought to be
responsible for reionization. Here we examine one such galaxy, Pox~186. 

\subsection{Pox~186}

Pox~186 was originally discovered as an emission line, Blue Compact Dwarf (BCD) galaxy in the objective prism survey of \citet{1981A&AS...44..229K}. The special character of Pox~186 was highlighted by \citet{1988A&A...204..10}, whose ground-based imaging showed no evidence of an underlying stellar population associated with the starburst and reported Pox~186 as a non-detection in a dedicated 
search with the Nancay radiotelescope for \hi\ 21cm emission.  These two characteristics, having a starburst with no apparent host galaxy and the presence of efficient star formation in the absence of a neutral gas ISM are exceptionally rare, if not unique.  We hypothesize that the current starburst in Pox~186 may have resulted in the complete blow-away of its neutral ISM. In this scenario, the ionized gas that we observe today is density bounded due to the lack of a surrounding neutral ISM, and, as a result, Pox~186, like a green pea galaxy, would be expected to have a relatively high escape fraction of ionizing photons.

Pox~186 has similar characteristics to Green Pea galaxies, in particular their density bound nature with the implication that they are optically thin to Lyman continuum radiation \citep{2013ApJ...766..91}.  Three observations point to Pox~186 being density bounded, where there is not enough material in the galaxy to absorb all ionizing radiation. If entirely density bounded, \fesc\ could reach close to 100\% and it would be the first case of a complete neutral gas blow-out observed in a galaxy. 

First, Pox 186 has an \ott $ > 20$ (\citealt{2004ApJ...421..519}, referenced as G04 from now on), one of the most extreme values among the Green Pea sample \citep{2017MNRAS...471..548}. This
implies a high ionization state in the majority of the ISM. 

Second, its S/O abundance ratio inferred from a standard nebular abundance analysis is $\sim$2 times
larger than the mean for BCDs (G04), which likely stems from an overestimated ionization correction factor
(ICF). This is most easily understood if the \hii\ region is density-bounded and inappropriate ionization
corrections derived from ionization-bounded \hii\ region models are used. Sulfur's ICF is
dependent on O/O$^+$, which is sensitive to the low ionization region at the edge of the \hii\ region as
O$^0$ has a similar ionization potential to neutral hydrogen. A density bounded nebula does not have the
layer of low ionization at the edge of the nebula. This decreases the abundance of O$^+$ relative to O
relative to a ionization bounded nebula, therefore increasing O/O$^+$ and causing an over-correction.

Finally Pox~186 remains undetected in \hi\ 21cm, despite attempts from multiple observatories (Nancay Radio Telescope, \citealt{1988A&A...204..10}; Giant Metrewave Radio Telescope, \citealt{2005MNRAS...362..609}; Very Large Array\footnote{John Cannon, Private Communication}).
With a 5$\sigma$ upper limit of M$_{HI} <  1.6$ $\times$ $10^6$ M$_{\odot}$, its M$_{HI}$/L$_{B}$ ratio is $\sim0.1$, much
lower than that for typical BCDs ($\sim1$), suggesting a large portion of the \hi\ has been ionized. 

In their modeling of feedback in dwarf galaxies, \citet{1999ApJ...513..142} distinguish between blow-out, when material is ejected from the galaxy's gravitational potential and blow-away, when the entire ISM is driven away from the galaxy.
If complete neutral-gas blow-away is indeed possible in a galaxy, one would expect the host galaxy to be low mass, and \citet{1999ApJ...513..142} found that only galaxies with gas masses $\le$ 10$^6$ M$_{\odot}$ were expected to be vulnerable to complete blow away. Observations and simulations of galactic outflows
support this, finding low mass galaxies form galactic outflows more efficiently due to a shallow potential well \citep{2017MNRAS...469..4831,2020MNRAS...494..3328}. 

Hosting a stellar mass of $\sim$10$^5$ M$_{\odot}$
\citep{2002ApJ...581..1039}, Pox~186 contains less mass in stars than typical BCDs 
or even individual star-forming clumps in massive star-forming galaxies \citep{2011ApJ...733..101}, and is more
comparable to giant \hii\ regions (e.g., NGC 604; \citealt{1996AJ...112..146}). It is extremely rare to observe a dwarf galaxy with stellar mass less than 10$^6$ M$_{\odot}$ hosting a starburst.  
Thus, Pox~186 falls in a mass regime where the effects
of feedback have rarely been studied observationally. Pox~186 is unique in this regard as it is simultaneously low mass, hosts a young super star cluster, and is nearby (D$\approx$13 Mpc) allowing for spatially resolved analyses.

Pox~186 exhibits a radial velocity of $\sim$ 1200 km s$^{-1}$. At this redshift, nearby mass concentrations such as the Virgo Cluster, the Great Attractor, and the Shapley Concentration cause significant deviations from the Hubble flow. Therefore, we use Cosmicflows-3 \citep{2017ApJ...850..207,2020AJ...159..67} to calculate a distance of 13.1 Mpc to Pox~186.
We adopt this distance and convert previous literature values using this distance for direct comparison.
The distance and other parameters of Pox~186 are collected and presented in Table 1.

\begin{table}
\centering
\begin{tabular}{lll}
\multicolumn{3}{c}{Pox 186 Basic Properties}     \\
\toprule
Parameter                  & Value      & Reference \\
\hline
RA (deg)                   & 201.452767        &   \\
DEC (deg)                  & -11.610494        &   \\
m$_V$ (mag)                & 17.43 $\pm$ 0.03  & 1 \\
log(M$_*$/M$_{\odot}$)     & 5                 & 2 \\
log(M$_{HI}$/M$_{\odot}$)  & $<$ 6.2           & 3 \\
SFR (M$_{\odot}$/yr)       & 0.045 $\pm$ 0.003 & 2 \\
\hb\ EW (\AA)              & 375.0 $\pm$ 0.6   & 1 \\
\oiii/\oii\ (\ott)         & 18.3 $\pm$ 0.11   & 1 \\
12 + log(O/H)              & 7.74 $\pm$ 0.01   & 1 \\
Distance (Mpc)             & 13.1              & 4,5\\ 
\hline

\end{tabular}
\caption{Properties of Pox~186. References are 1 \citep{2004ApJ...421..519}, 2 \citep{2002ApJ...581..1039}, 3 \citep{2005MNRAS...362..609}, 4 \citep{2017ApJ...850..207} , and 5 \citep{2020AJ...159..67}.}
\end{table}

In this paper we present new Gemini Integral Field Spectroscopy (IFS) data to measure kinematic properties of the
ionized gas across Pox 186 and explore the different mechanisms that can drive the observed nebular
emission. Section \ref{sec:data} describes the data and reduction process while the results are shown in
Section \ref{sec:results}. The discussion and conclusions are presented in Sections \ref{sec:discussion} and
\ref{sec:conclusions}.

\section{Data and Methods} \label{sec:data}
\subsection{Observations}\label{sec:obs}
\begin{figure}[ht!]
\plotone{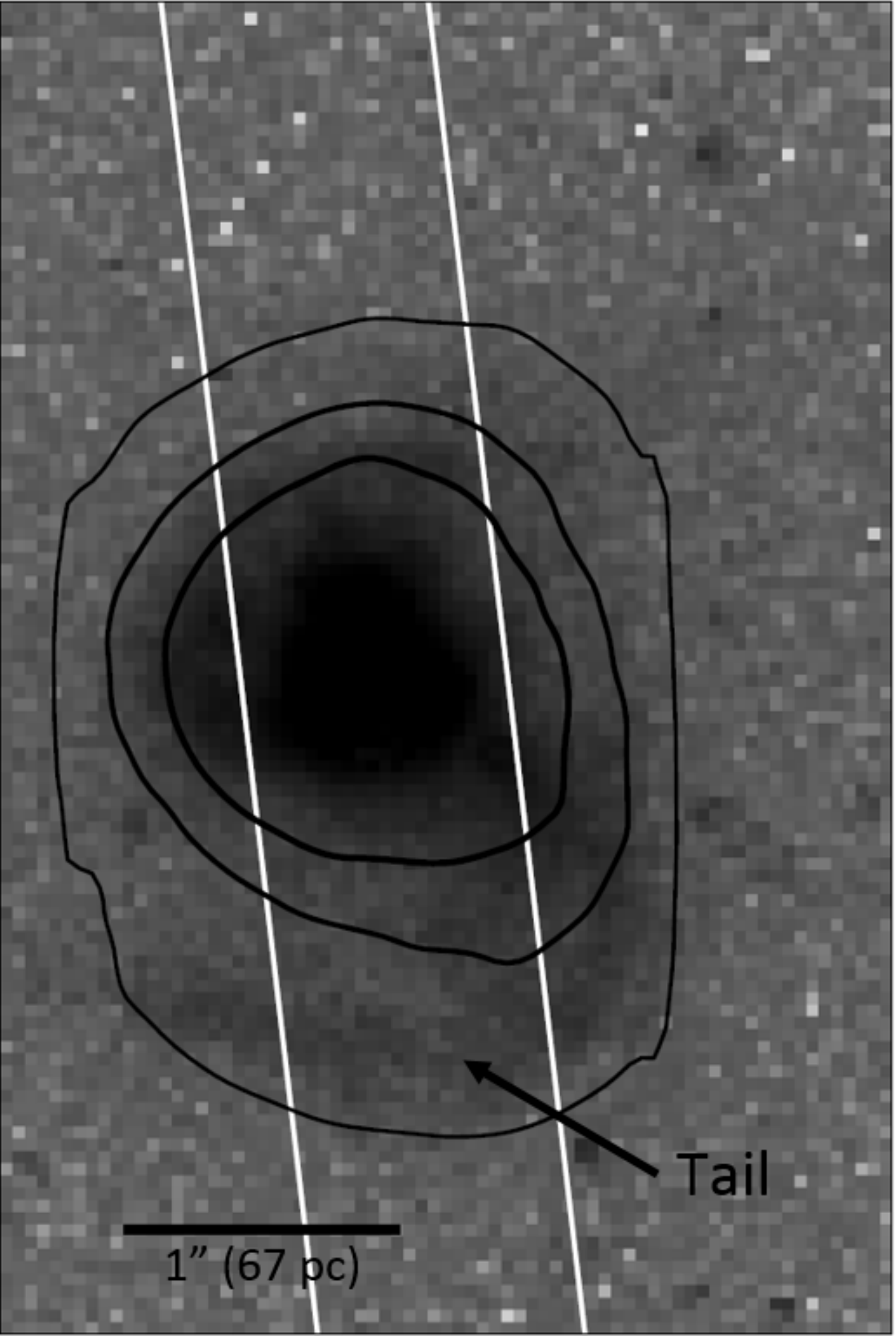}
\caption{HST F555W image of Pox 186. The black contours denote \oiii\ S/N ratios of 24, 25, and 26 from the
Gemini data cube while the white lines show how the slit from G04 lines up with the field of view of the data cube. The horizontal black line represents
1\arcsec. For reference 1\arcsec\ corresponds to a linear scale of 67 pc. \label{fig:HST}}
\end{figure}

\noindent
Data were taken with Gemini-South using the Gemini Multi Object Spectrograph (GMOS) IFU over five, 1000
second exposures on April 14, 2018. We used the one slit mode, with a field of view of 3\arcsec $\times$
5\arcsec\ (corresponding to a physical scale of 196 pc $\times$ 325 pc; Figure \ref{fig:HST}), and the B600 grating for an effective wavelength range from 3650 \AA\ to
6600 \AA. The spectral resolution was $\sim1.6$ \AA\ (FWHM) and seeing was $0.\arcsec88$. Standard star
LTT3864 was observed for flux calibration and spectra of a CuAr comparison lamp for wavelength calibration.

The data are reduced primarily using Gemini's provided IRAF software package. The package includes bias and
overscan subtraction, flat fielding, cosmic ray removal, wavelength calibration, scattered light and quantum
efficiency corrections, fiber tracing and extraction, sky subtraction, flux calibration, and mapping each
fiber into the data-cube. Due to the unreliability of the overscan and scattered light modeling routines,
custom scripts in python were written and used instead. Bad columns were masked by hand before the reduction
process, and the final data-cube products of each exposure were aligned by cross-correlating the images of
the brightest nebular lines. The aligned data-cubes were then combined using a custom script written in
python. Finally, a foreground Galactic extinction correction of $E(B-V)=0.0385$ was applied to all spectra using the
optical extinction curve from \citet{2011ApJ...737..103} assuming an $R_V$ value of 3.1. 
This very low correction for foreground Galactic extinction is expected due to the high 
Galactic latitude of $+50.4^\circ$ for Pox~186. 
\begin{figure*}
\plotone{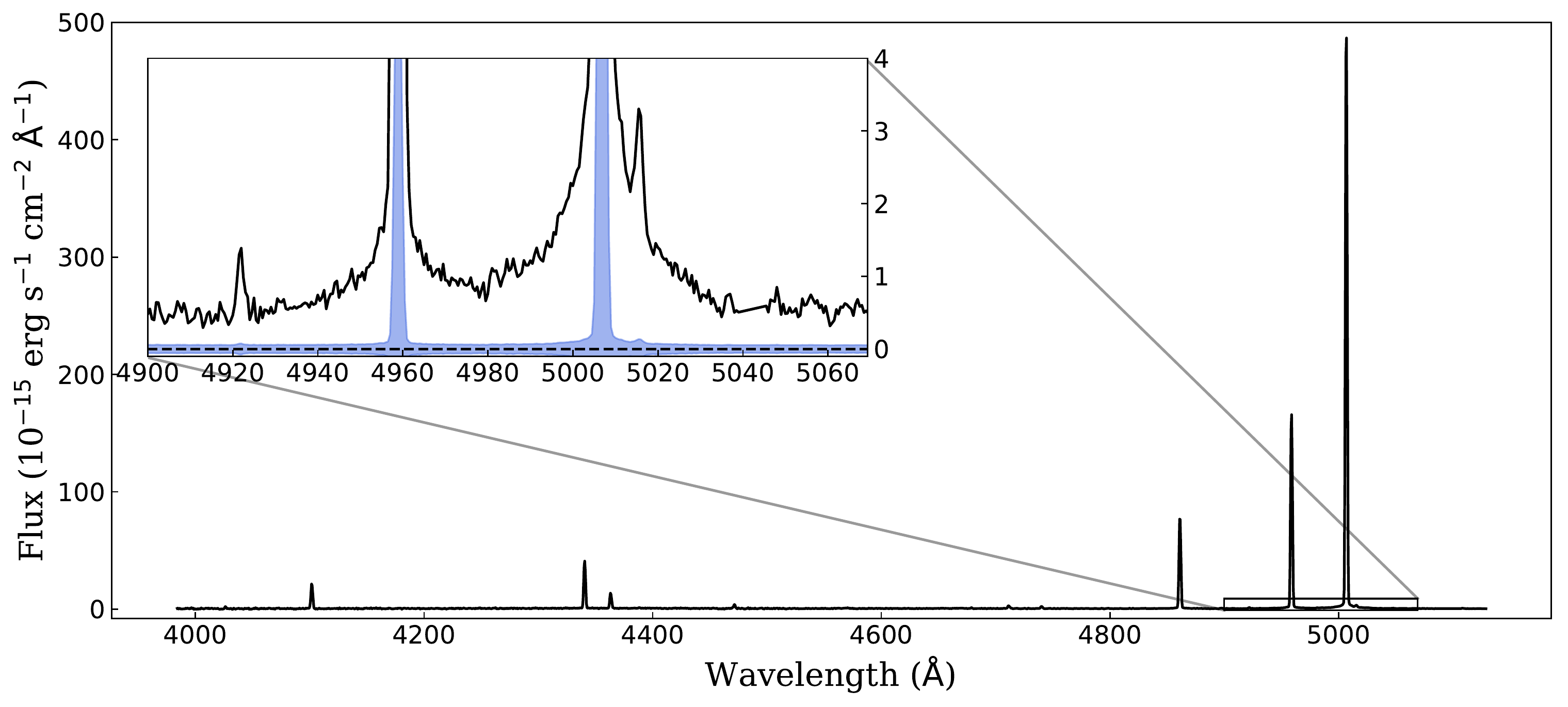}
\caption{One dimensional spectrum. The black line is the spectrum while the blue shaded region represents
the 1$\sigma$ errors including the additional 3.8\% of the flux as mentioned in Section \ref{sec:data}. The
zoomed in region shows the very broad wings present on the \oiii\ emission lines.}
\label{fig:1dspec}
\end{figure*}

The CCD array used to take the data is an array of 3 chips, each with 4 amplifiers for 12 amplifiers total.
The chip covering the red part of the spectrum experienced a charge transfer inefficiency problem for a time
that included when the data were taken. This problem made data beyond $\lambda > 5450$ \AA\ unusable.
Therefore, we do not consider any lines beyond $\lambda > 5450$ \AA. In addition, variation between
exposures is larger than the calculated errors provided by the pipeline. Comparing line fluxes across the 5
different exposures we find a 1$\sigma$ scatter of $\sim3.8\%$. We therefore increase the error estimate by an amount of $3.8\%$ of the flux to account for this scatter. 

To be free from the uncertainty of the atmospheric transmission during the standard star observations, we calibrate our data to HST archival 
Wide Field Planetary Camera~2 F555W observations \citep[HST-GO-8333, PI $=$ Corbin;][]{2002ApJ...581..1039}. We simulate a F555W observation using the IFU data by calculating the mean flux density over the F555W bandpass
using 
\begin{equation}
    f_{\lambda}(P) = \frac{\int P(\lambda)f_{\lambda}(\lambda)\lambda d\lambda}{\int P(\lambda)\lambda d\lambda}
\end{equation} where $f_{\lambda}(\lambda)$ is the flux density  and $P(\lambda)$ is the filter throughput
at each wavelength. We find a factor of 3 discrepancy between our standard star flux calibrated data and HST photometry. After correcting for this discrepancy, we then compare the measured line fluxes to G04, finding good agreement within the errors.

Balmer ratios were examined to correct for the extinction from the host galaxy. As \ha\ fell on the CCD chip with the charge transfer inefficiency, it was not considered for this correction. We find both \hd/\hb\ and \hg/\hb\ to be consistent with Case B within the uncertainties, 0.254$\pm$0.014 and 0.488$\pm$0.025 (as compared to the theoretical values of 0.263, and 0.474) respectively. 

This inferred very low extinction is expected for a low-metallicity galaxy, but the literature values for the extinction to Pox~186 show some inconsistencies. 
The original spectroscopy of \citet{1981A&AS...44..229K} reported an H$\alpha$/H$\beta$ ratio of 3.3, which converts to an E(B-V) of 0.15, and \citet{1983ApJ...273...81K} derived a value of  c(H$\beta$) $=$ 0.39, which translates to an E(B-V) of 0.27.
\citet{2002ApJ...581..1039} report a value of E(B-V) $=$ 0.28, in agreement with the larger of the two values.
From two independent measurements, \citet{2004ApJ...421..519} report lower values of
C(H$\beta$) $=$ 0.02 and 0.16 (E(B-V) equivalents of 0.014 and 0.11, respectively).
Although the very high value of  \citet{2002ApJ...581..1039} does not take into account underlying stellar absorption and that the H$\alpha$ emission is blended
with [N~II] emission, and both of which would tend to overestimate the extinction, both of these effects are estimated to be at the level of 1\% in flux, and so are unlikely the cause for the discrepant high value of extinction.  Because our derived value for extinction is consistent with zero, we expect a low value of extinction due to the low metallicity of Pox~186, and our measurement agrees with the value obtained with the MMT by \citet{2004ApJ...421..519}, we do not perform a correction for internal extinction for Pox~186.  This choice has essentially no impact on any of our later conclusions.

Using 19 emission lines with the highest signal to noise we derive a spectroscopic redshift of $z=0.00409 \pm 0.00006$, consistent with G04, who measure a redshift of $z=0.00413 \pm 0.00005$. 

\subsection{Kinematic Component Fitting} \label{sec:Datafitting}

Broad wings are prevalent in the strongest emission lines (\oiii\ and \hb; Figure \ref{fig:1dspec}), suggesting multiple kinematic
components may be present in the ionized gas of Pox 186. Spectroscopic lines from the CuAr comparison lamp are well described by a single Gaussian (S/N $>$ 50) and therefore a nonlinear component in the instrument response cannot be responsible for the broad wings. To quantify the wings (also referred to as the ``broad component"), we fit a multi-component emission
line model to the \oiii\ doublet using Markov Chain Monte Carlo (MCMC) fitting. 

We use the python implementation emcee \citep{2013PASP...125..306} for our MCMC fitting using 100 walkers
and iterating over 5000 steps, with the first 500 as the burn in phase. To ensure our continuum measurement
is not being contaminated by the broad component we measure the continuum from $4780-4840$ \AA. Our model
uses uniform priors spanning all possible values for the parameters. \oiii\ 4959 \AA\ and 5007 \AA\ are fit
simultaneously by keeping the line center offsets and Gaussian widths consistent between the two while
scaling the amplitude by their intrinsic ratio, 2.94 \citep{2006USB...84}.

The narrow component is well described by a Gaussian distribution while the wings are not. 
First, to characterize the width of the wings in a self-consistent manner, we fit a two Gaussian model using
the MCMC routine described above. After each spatial pixel (spaxel) is fit, the broad component flux S/N is
evaluated as a proxy for the quality of the overall fit. The uncertainties derived from the distribution of
walkers from the MCMC fit are used to measure the S/N of the flux. Spaxels with a broad fit with a S/N $<$ 3
are then refit with a single Gaussian model. Note that we mask out the faint \hei\ (5016 \AA) line blended with
the wing of 5007 \AA\ and \hei\ (4922 \AA) during the likelihood maximization.

Velocity dispersions of each kinematic component were corrected for thermal broadening and the instrument
response, $\sigma_{int}=\sqrt{\sigma_{G}^2 - \sigma_{T}^2 - \sigma_{R}^2}$, where $\sigma_{G}$ is the fit
dispersion of one of the Gaussian components, $\sigma_{T}$ is the thermal broadening, and $\sigma_{R}$ is
the velocity width of a spectral resolution element. Thermal broadening was measured from the electron
temperature assuming a Maxwell-Boltzmann distribution: $\sigma_T=\sqrt{3kT_e/m}$ where $m$ is the mass of a
doubly ionized oxygen atom and T$_e$ is the electron temperature measured with the direct method (\oiii\
4363 \AA\ and \oiii\ 4595 and 5007 \AA) using equation 5.4 from \citet{2006USB...84}. Electron
temperatures derived from \oiii\ 4363 \AA\ weakly depend on the electron density. As we do not detect \oii,
the data exclude the \sii\ doublet, and the \ariv\ doublet traces more highly ionized gas than \oiii, we
assume an electron density of 350 cm$^{-3}$ as measured by G04. Summing all spaxels into one spectrum we
measure T$_e = $ 16780 $\pm$ 470 K, while the average T$_e$ across spaxels with a 3$\sigma$ detection of
\oiii\ (4363\AA) is 16920 K with an RMS scatter of 450 K.  For reference G04 measure a temperature of $16940
\pm 60$ K.  Finally, the spectral resolution, $\sigma_R$, is measured empirically from the arc lamp exposures
to be 41 \kmps.

\section{Ionized Gas Kinematics in Pox~186} \label{sec:results}

\begin{figure*}[ht!]
\plotone{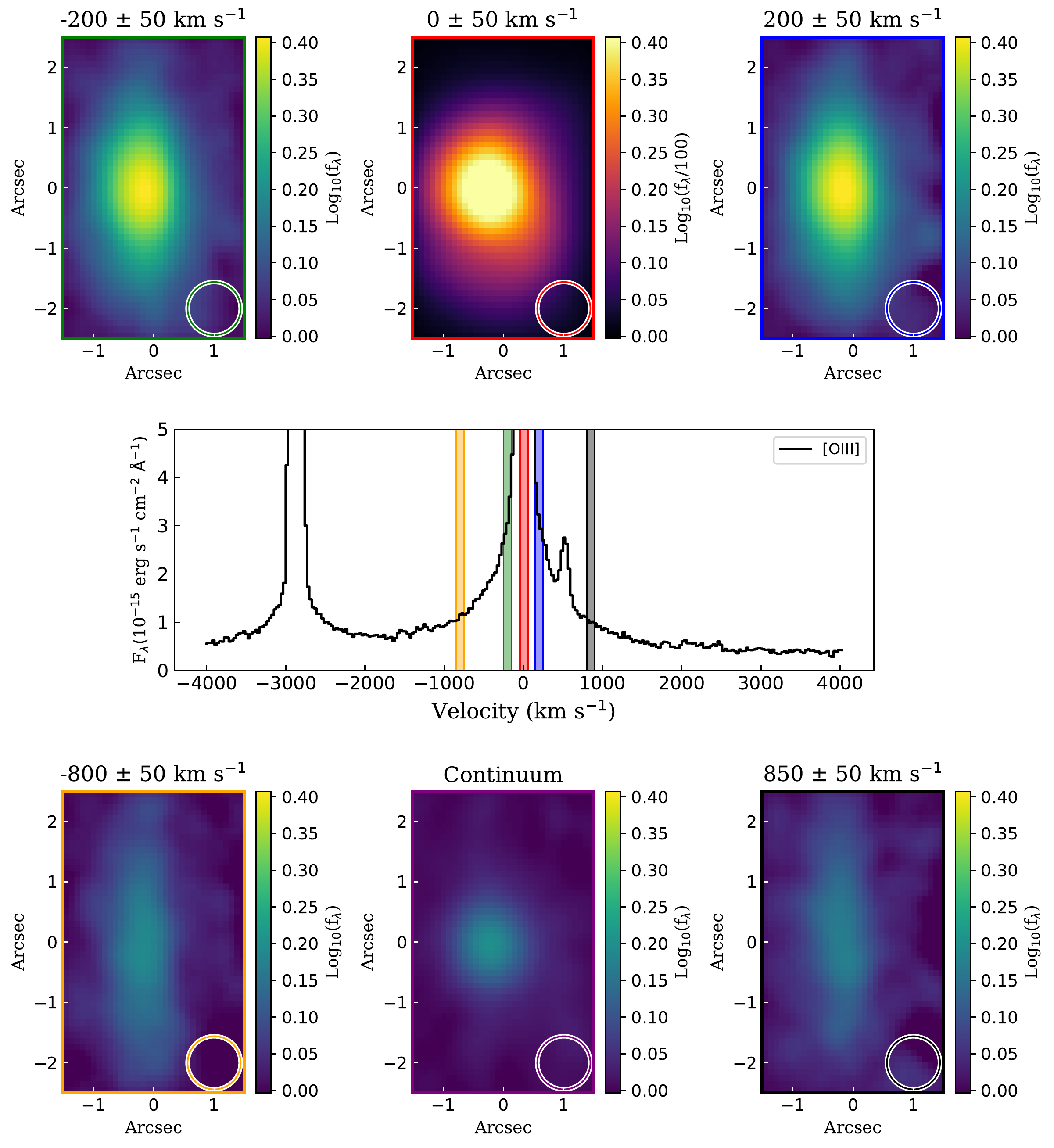}
\caption{Flux distributions from gas at different velocities. The central panel shows the 1-dimensional
spectrum summed over all spaxels while the shaded bars signify the regions in velocity space used to create
the surrounding images. Each outer image has been continuum subtracted except for the continuum image. The
colors of the bars match the outline of their corresponding image as well as the circle visible in the
bottom right corner of each. The diameter of the circle denotes the full width half max of the seeing,
.88\arcsec. The upper central panel has been scaled by 100. \label{fig:velslices}}
\end{figure*}

We find the broad component is not detected in all spaxels. Indeed, mapping \oiii\ according to its radial
velocity shows a transformation in geometry (Figure \ref{fig:velslices}). At the line center (red shaded
region and upper-middle image), the emitting gas has a circular geometry, similar to the continuum image but
with an extended ``halo" around the bright core. This halo is likely the low surface brightness component identified in G04
detected out to large radii (r$\sim$6\arcsec) in ionized gas. Like G04, we find no evidence of stellar continuum at larger radii. The
surrounding radial velocity slices of Figure \ref{fig:velslices} highlight the transformation from a
circular geometry to elongated at high radial velocities. 

\begin{figure*}
\plotone{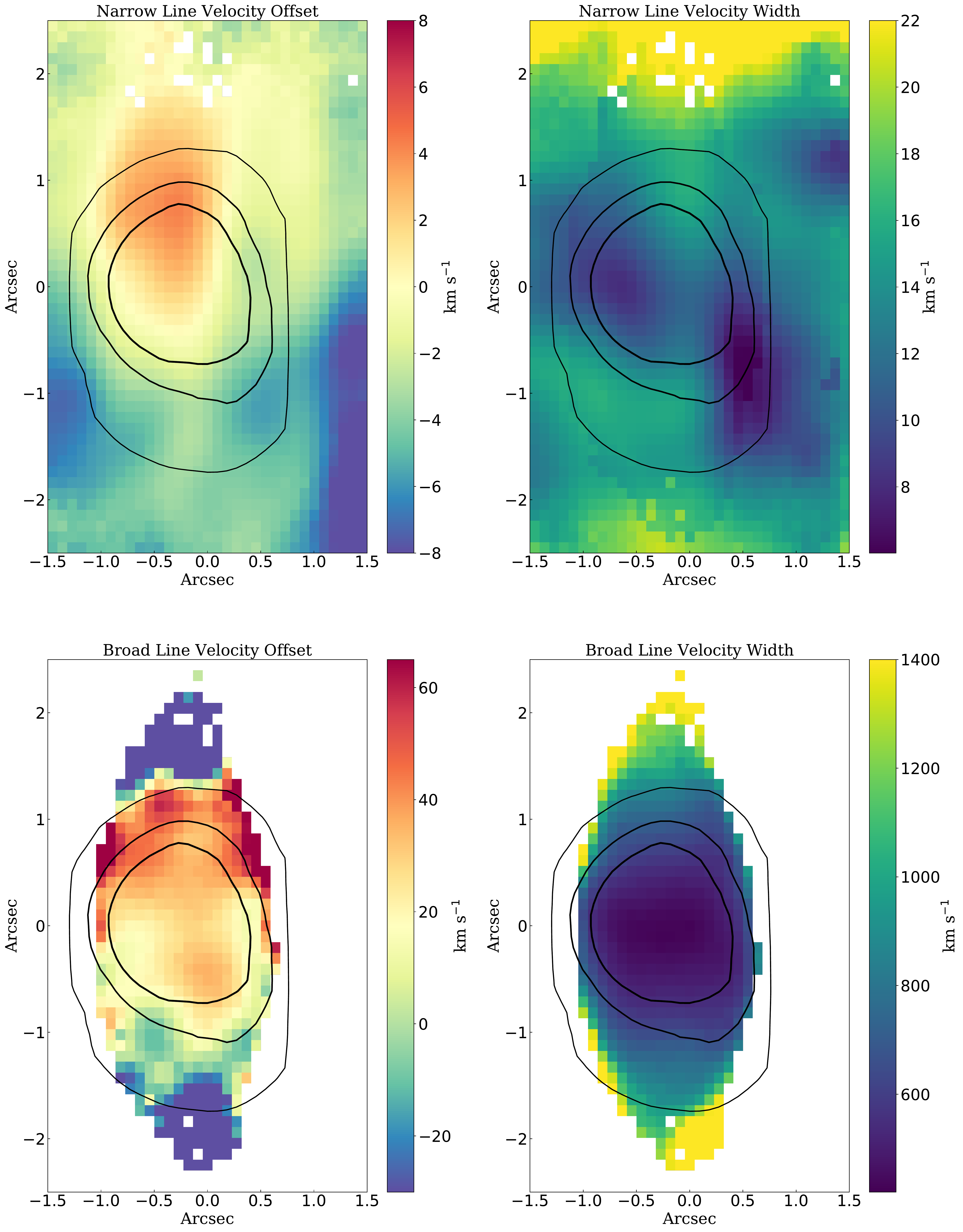}
\caption{Kinematics of the two Gaussian components as fit by two-Gaussian model described in Section
\ref{sec:data}. Velocity dispersions are corrected for thermal dispersion and instrument response. Black
contours are the same as shown in Figure \ref{fig:HST}.}
\label{fig:pars}
\end{figure*}

This elongated geometry of the high velocity gas is seen in the results of the MCMC fits, shown in Figure
\ref{fig:pars}. Spaxels where the S/N in the component's flux is less than 3 are cut out of the maps. This cut is relevant only for the broad component as the narrow component is detected
with a S/N $>$ 3 in all spaxels. The black contours are the same as in Figure \ref{fig:HST}, corresponding
to S/N contours in total \oiii\ flux of 24, 25, and 26, and are included to help guide the eye. 
Table \ref{tab:errors} list the general properties of the uncertainties in the parameters shown in
Figure \ref{fig:pars}. The second column lists the median uncertainty, and the third column lists the
1-$\sigma$ width of the distribution of uncertainties of the spaxels shown ($\sigma_{err}$). 

\begin{table}
\centering
\begin{tabular}{lcrr}
\hline
MCMC Fit Parameter & Median Uncertainty & $\sigma_{err}$ \\
\hline
Narrow Velocity Offset & 1.05 & 1.85\\
Narrow Velocity Width & 0.84 & 1.85\\
Broad Velocity Offset & 92.9 & 89.3\\
Broad Velocity Width & 98.5 & 223.9\\
\hline
\end{tabular}
\caption{Median error and 1$\sigma$ width of the distribution of errors for the each fit parameter. All values listed are in units of \kmps.}
\label{tab:errors}
\end{table}

The top two panels of Figure \ref{fig:pars} show maps of the velocity offset and velocity width of the
narrow component. Velocity offsets are measured as $v_{off}=c\Delta\lambda/\lambda_0$, where c is the speed
of light, $\Delta\lambda=\lambda_f-\lambda_0$, and $\lambda_0=5006.84(1+z)$ while the velocity widths
($v_w$) are measured as $v_w=c\sigma_f/\lambda_0$, where $\sigma_f$ is the standard deviation of the
Gaussian component. In the upper half of the field of view of the velocity offset map, there is a feature of
the narrow component with a small ($\sim6$ \kmps) systematic velocity shift, indicating a flow of gas that
appears to originate in the core of the galaxy. There also seems to be a small gradient between the upper
and lower halves of the field of view that could be evidence of rotation. The amplitude of the gradient is $\sim$2 \kmps\ and is consistent with what one would expect of a galaxy the size of Pox~186, but is also comparable to the mean uncertainty in the velocity offsets (Table
\ref{tab:errors}). 

The velocity width map of the narrow component shows a small enhancement ($\sim17$ \kmps) in the tail region
while the top and bottom of the field of view show larger enhancements ($\sim22$ \kmps) relative to the
central region of the galaxy ($\sim11$ \kmps). 

Similarly, velocity offsets and widths of the broad component are shown in the bottom two panels of Figure
\ref{fig:pars}. A region of gas with a systematic velocity shift similar to the narrow component is also
present in the broad component velocity offset map. The projected spatial location and amplitude of the
offset are slightly different from the narrow component, but the map is limited in its usefulness due to the
offset uncertainties. The offset uncertainty properties of the broad component in Table \ref{tab:errors}
highlight that outside the central 0.5\arcsec\ of the galaxy, the fit offset of the broad component is
consistent with 0. 

Unlike the broad offsets, the increase in width at larger radii is not a function of the fit uncertainty.
The uncertainties in kinematic parameter fits are shown in the Appendix. Figure \ref{fig:mcmcerrs} shows the
uncertainties follow a different distribution than the fit broad velocity widths. The broad component width
increases smoothly with increasing distance from the core of the galaxy, with an asymmetric gradient where
the width increases faster in the vertical directions than horizontal. Conversely the uncertainties of the
broad component widths are consistently small except for the very edges of the wing region, due to the
decrease in the S/N of the broad component. In addition, there is a similar level of correspondence between
the narrow and broad velocity widths as there is in the offset maps. While there is no observed enhancement
in the tail region that passes the S/N cut, the general increase in both component widths at increasing
radii suggests some level of interaction between the narrow and broad component gas. 

\begin{figure}[ht!]
\plotone{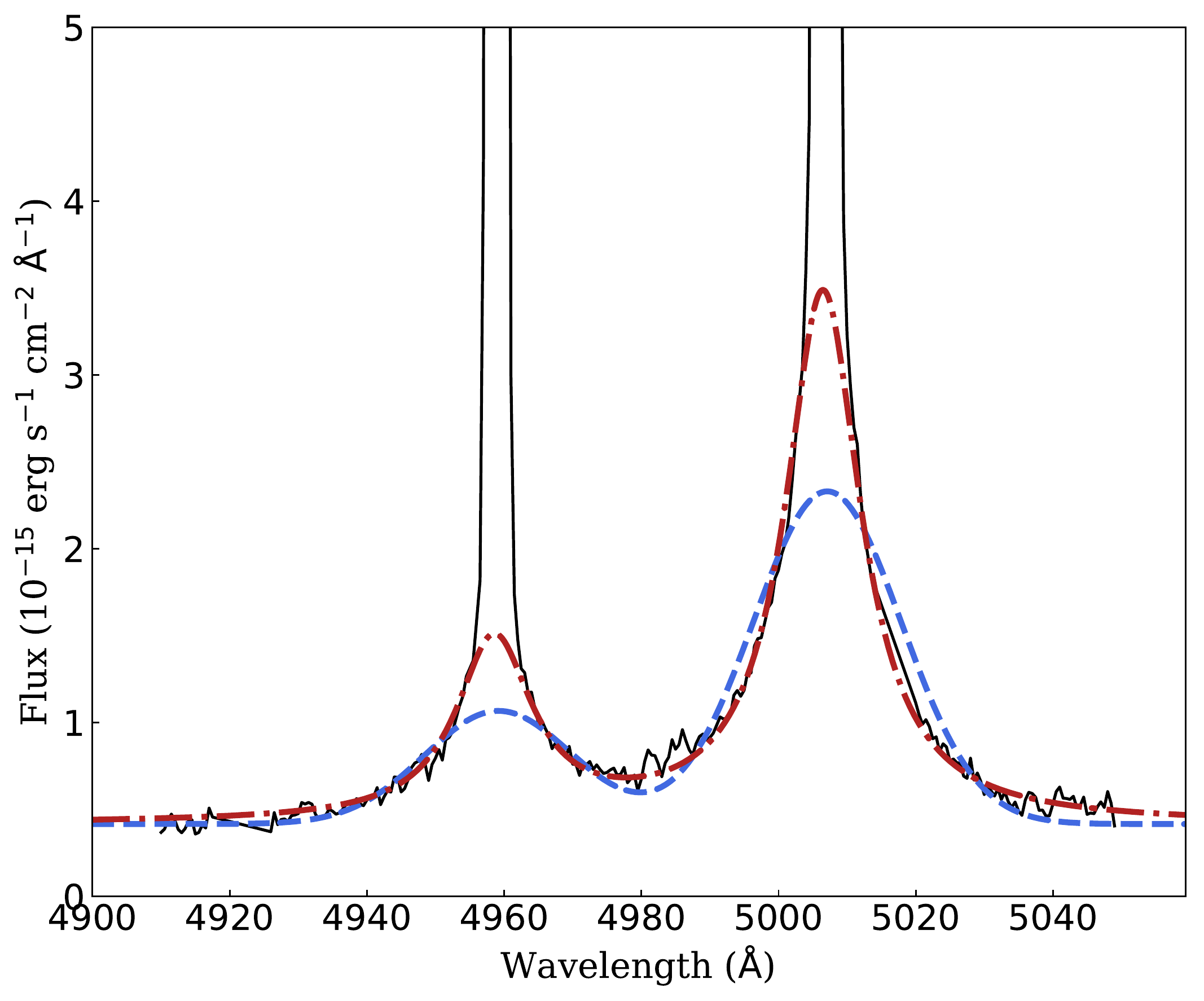}
\caption{Broad component of \oiii\ in the wing region. The solid black line is the data, while the dashed
blue line is the Gaussian fit, and the dot-dashed red line is the Lorentzian fit. A Lorentzian profile is
significantly better at describing the broad component line shape and isn't as prone to underestimating the
broad component flux as the Gaussian fit near the core of the line. \label{fig:residual}}
\end{figure}

We find the wings are better fit by a Lorentzian profile convolved with the instrument response as compared
to a Gaussian (Figure \ref{fig:residual}). The Lorentz profile fit to the wings of \oiii\ have a FWHM of 884
\kmps\ and account for 7.6 $\pm$ 0.1\% of the total flux of the line. Even binning all spaxels where the broad component is detected in \oiii, the broad component of \hb\ has a large uncertainty, accounting for 6.5 $\pm$ 3.8\% of the total line flux and is slightly
narrower, with a FWHM of 760 \kmps.

\section{Outflow Properties} \label{sec:discussion}

\subsection{Outflow Rates}
Ionized gas emission is observed at extremely large velocities ($|$v$|$ $> 1000$ \kmps) but gas at large
velocities does not necessarily imply an outflow. To determine if the high velocity gas constitutes an
outflow, the escape velocity ($v_{esc}$) is estimated following \citet{1995ApJ...438..563}, who assume a
spherically symmetric isothermal potential and a cutoff at $r_{max}$. The escape velocity at radius r becomes:
\begin{equation}
    v_{esc}=\sqrt{2}v_{circ}\sqrt{1+ln(r_{max}/r)} \label{eq:2}
\end{equation}
where, for Pox~186, $r_{max}$ is taken to be 100 pc (1.5\arcsec), and $r$ to be 1/2 $r_{max}$. $v_{circ}$ is the circular
orbital velocity estimated by the virial theorem: $v_{circ}=\sqrt{GM_{tot}/r_{max}}$.  The total mass is
estimated by assuming a dark matter fraction of 90\%, making M$_{tot}$=M$_{baryon}$+M$_{DM}$. M$_{baryon}$ is the
baryonic mass component and consists of stellar mass, molecular, neutral, and ionized gas masses.  

The ionized (\hii) mass is given by: 
\begin{equation}
    M_{H}=\frac{\mu m_H \lambda_{H\beta}L_{H\beta}}{hc\alpha^{eff}n_e} \label{eq:3}
\end{equation}
where $\mu$ is the atomic weight, $m_H$ is the mass of hydrogen, $\lambda_{H\beta}$ is the wavelength of
\hb, $L_{H\beta}$ is the luminosity of the \hb\ line,  $h$ is Planck's constant, $c$ is the speed of light,
and $\alpha^{eff}$ is the recombination coefficient for \hb. We take $\mu$ to be 1.36 to account for an
assumed 10\% He fraction, and $\alpha^{eff}$ to be 1.61$\times$10$^{-14}$ cm$^3$ s$^{-1}$ assuming case B
recombination in 20000 K gas \citep{2006USB...84}. Using this equation we measure an \hii\ mass of $5.4\ \pm\
0.38 \times 10^4$ M$_{\odot}$. We then take the stellar population synthesis model from \citet{2002ApJ...581..1039} which found the central star cluster to have a M$_{*} \sim 10^5$ M$_{\odot}$, adjust the limit  on the \hi\ mass from \citet{2005MNRAS...362..609} for the cosimicflows-3 distance to M$_{HI}$ $<$
$8.0\times10^{5}$ M$_{\odot}$, and assume the neutral
and molecular masses to be comparable to the neutral hydrogen mass upper limit, $\sim8 \times 10^5$ M$_{\odot}$. With
these assumptions we estimate the upper limit on the total mass of Pox~186 to be $1.8\times 10^7$ M$_{\odot}$
and on the escape velocity to be $\sim 55$ \kmps. The ionized gas mass being comparable to the stellar mass is significant, and could be further evidence for a possible blow-out of the neutral gas. It is also worth noting the assumptions made above are all made
conservatively to maximize the total mass and therefore escape velocity, making these values upper limits.
Considering the radial velocities observed in the ionized gas are an order of magnitude greater than the
upper limit on the escape velocity, it is very likely that most of the high velocity wind is not bound to
Pox~186 and will escape into the IGM.

Next, the mass outflow rate of the ionized phase of the outflow is estimated. The
ionized gas mass of the outflow can be measured from the total flux in the broad component assuming the
entire component is high velocity gas that will escape the galaxy. Spaxels where the broad wings are
statistically significant (S/N $> 3$) are summed into one spectrum to provide the highest possible S/N in
the wings. A Lorentzian profile is then fit to the wings finding the broad component contributes 6.5 $\pm$
3.8 \% to the total \hb\ line flux, which we refer to as the ``flux fraction".
Using equation 2 gives an ionized gas mass of 6.4$\pm$3.7 $\times$ 10$^{3}$ M$_{\odot}$ in the broad
component. Equation 2 can be modified to find the mass outflow rate of the wind by multiplying by the flux
fraction ($F_{broad}/F_{line}$) and dividing by the time it takes the wind to move through the length of
the outflow ($R_{out}/V_{out}$):
\begin{equation}
    \dot{M}=\frac{\mu m_H \lambda_{H\beta}}{hc\alpha^{eff}n_e} L_{H\beta}\frac{F_{broad}}{F_{line}}. \frac{V_{out}}{R_{out}}
\end{equation} 
Assuming the characteristic length of the outflow ($R_{out}$) is 100pc (1.5\arcsec), the outflow velocity is
450 \kmps\ ($V_{out}$), and the flux fraction is 0.065, we calculate a mass outflow rate of 0.016 $\pm$ 0.010
M$_{\odot}$ yr$^{-1}$. It is worth noting the dominant source of error is the uncertainty in the
flux fraction from the fitting routine. If instead of using \hb\ we use the brighter \oiii (5007 \AA) flux fraction we calculate
$\dot{M}$=0.019$\pm$0.001 M$_{\odot}$ yr$^{-1}$. 

\subsection{Mass-Loading Factor}
The mass-loading factor measures  the efficiency of the outflow by normalizing the mass outflow rate
by the star formation rate (SFR). We follow the description in \citet{1998ApJ...498..541} and correct it to use the \hb\ luminosity to derive the SFR in Pox~186. 
Doing so, we find a SFR of 0.024 $\pm$ 0.003, corresponding to   a mass-loading factor of
0.67 $\pm$ 0.42. As before, the dominant source of uncertainty results from the fitting routine. Adopting the \oiii\ flux fraction and uncertainty results in a mass-loading factor of 0.8 $\pm$ 0.1. We emphasize the numbers derived above are estimates. Outflows are multi-phase phenomena, and the
assumption that the entire broad component is high velocity gas that will escape the galaxy is not necessarily entirely appropriate.

\subsection{Discussion on the Line-Shape of the Broad Component}
Outflow geometry is typically invoked to explain the line-shape of the emission lines tracing it. We consider
reasonable geometries that conform to the line-shape and spatial extent of the broad component. From the spatial extent (Figure \ref{fig:velslices}) the outflow seems to have a bi-conical or ``disk-like" shape. If a radial outflow from the central star cluster is considered, then a bi-conical outflow seems to be the favored geometry. In the core of the galaxy, the width of the broad component is smallest and increases at larger radii, the opposite of what would be expected from a ``disk-like" geometry where the core would show the broadest wings from seeing both gas components moving directly along the line of sight. The broad velocity offsets imply the upper bi-cone is angled away from the observer slightly while the lower bi-cone is consistent with being perpendicular to our line of sight. Increasing widths of the broad component at larger radii, however, are incompatible with both geometries assuming a simple radial outflow. 

The gas in the broad component is likely not all traveling at the same speed and has a velocity distribution
of its own in addition to how the velocities line up with the line of sight. Considering both effects, it is
surprising that the line-shapes of the wings are so consistent across the entire wing region, which spans 4 spatially resolved beams along the vertical axis in Figure \ref{fig:pars}. If the Lorentzian profile is
just a coincidence of geometry of the outflow, then it would be extremely unlikely to have the same
line-shape across the whole wing region which is larger than the seeing and spans above and below the
central starburst. Therefore, it seems likely that the mechanism behind the wing's line-shape needs to be a
general phenomenon and should be independent of chance alignments of filaments of gas.  

The effects of the outflow on the ionized gas kinematics can be seen in the velocity offset and width maps
(Figure~\ref{fig:pars}). Both velocity offset maps have a similar geometry with a redshifted component in the
upper half of the field of view, but with different amplitudes, $\sim$4 \kmps\ and $\sim$40 \kmps\ for the
narrow and broad respectively. In the context of an outflow, consistency in radial velocities could be due to the
entrainment of cooler gas clouds by the higher velocity wind. As the wind moves outward, it interacts
with the stationary, denser gas clouds by imparting some of its momentum to the stationary gas,
effectively sweeping up the clouds into the flow. In this scenario, the motions of the denser clouds are not
dominated by virial motion in the gravitational potential well. The velocity offset maps support this
further as they do not show strong evidence of rotation. More likely, the starburst itself and the feedback
it produces are the primary drivers of the kinematics seen in Pox~186. 

We are unable to explain the enhancement in the narrow component velocity dispersion at the top and bottom
of the field of view. Increasing uncertainties in the fit can partly explain the enhanced widths, but not completely. It is interesting, however, that this enhancement is seen at the “ends” of the
elongated region where the broad component is detected (Figure \ref{fig:velslices}). In the context of
condensations of denser gas being blown out, this enhancement could be the eventual “evaporation” or
break-up of the dense cloud as it disperses into the low surface brightness component. This low surface
brightness component could be indicative of ionized gas ejected from the galaxy, escape of ionizing
radiation to large radii, or both. Unfortunately, we are unable to map the full spatial extent of the
enhanced velocity dispersion around Pox~186.

We use the summed spectrum of the ``wing region" to explore the differences in the state of the high and low
velocity gas by comparing the \oiii(5007+4959)/\hb\ ratio of the narrow component and the wing component
separately. \oiii/\hb\ is affected by the ionization parameter, metallicity, and density of the emitting gas
making it particularly useful to qualitatively compare different gas components. 
We adopt the same fitting technique described in Section \ref{sec:Datafitting} and use a Lorentz profile
convolved with the instrument response for the wings and a Gaussian in the core. The \oiii/\hb\ ratios are
9.7 $\pm$ 5.7 in the wings and 8.3 $\pm$ 0.6 in the core. While these two ratios are consistent with one
another at a 1$\sigma$ level, with deeper data it is possible this discrepancy could be proven real. If so,
it could provide insights into the mechanism behind the broad component such as an enhanced collision rate, higher ionization state,
or be indicative of a metal enriched outflow. 

\subsection{Outflows in Low-Mass Galaxies}

Catastrophic cooling has been proposed as a mechanism for creating the high \ott\ ratios and large
\oiii\ luminosities seen in GPs \citep{2004ApJ...610..226,2017ApJL...849..L1,2019ApJ...885..96}.  
In low metallicity environments, the mechanical
feedback from star formation is suppressed which allows for dense ionized gas clouds to stay closer to the
star cluster without being driven outward and merged together. This is in contrast to when the dense clouds
are blown out where they effectively create a shell around the starburst. These lower metallicity, denser
clouds close to the starburst are extremely efficient at cooling (through \oiii\ primarily), and a side 
effect of not being blown out is that the high ionization, low density channels between clouds are preserved. 
Ionizing radiation and outflows driven by stellar processes escape through these channels.

At the interface between the low density channels and the higher density clouds, interactions between the
two phases are expected. \citet{2020arXiv...2002.10468} showed this interaction is expected to transfer momentum
from the low density winds to the high density clouds in the form of mixing. Turbulent mixing layers have
commonly been proposed as an alternative mechanism for broad emission lines but it is especially attractive
here due to the relative strengths of the two gas components and their homogeneity across the galaxy. In 
addition, the growth in velocity width of the broad component with increasing radii is consistent with 
entrainment of the relatively cool (T$\sim$10$^4$ K) gas into a fast moving, hot wind \citep{2001Experiments...30..65}.

\begin{figure*}[ht!]
\plotone{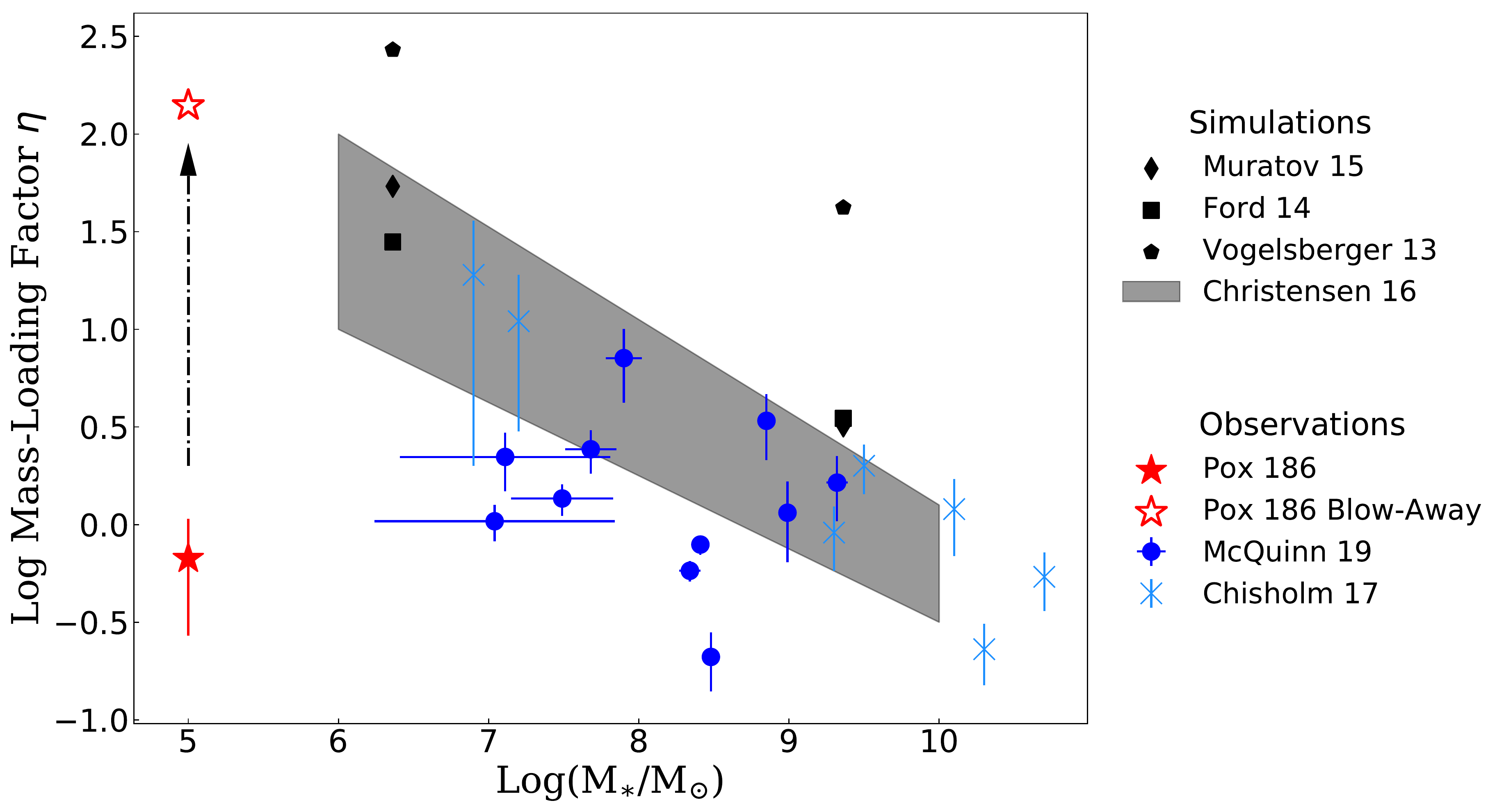}
\caption{Mass-loading factors as a function of stellar mass. Both the observed mass-loading factor of Pox~186 
(filled red star) and the mass-loading factor necessary to remove the expected \hi\ mass from galaxy scaling 
relations (empty red star) are shown. For context, we include observations from 
\citet{2019ApJ...886..74} shown with blue circles, and \citet{2017MNRAS...469..4831} shown in light blue 
crosses. Simulation predictions are shown with the black diamonds \citep{2015MNRAS...454..2691}, squares 
\citep{2014MNRAS...444..1260}, and pentagons \citep{2013MNRAS...436..3031}. We also show the range of predictions 
from \citet{2016ApJ...824..57} with the grey shaded region.  \label{fig:massload}}
\end{figure*}

Figure \ref{fig:massload} compares the calculated mass loading factor for Pox~186 with observations of 
other dwarf starburst galaxies and simulations. It is typically thought that low mass starburst galaxies 
tend to have higher mass-loading factors as their low mass makes it easier to blow gas out of their 
gravitational potential wells. Simulations (shown in black and grey in Figure \ref{fig:massload}) support 
this view \citep{2013MNRAS...436..3031,2014MNRAS...444..1260,2015MNRAS...454..2691}. Observations of 
mass-loading factors, however, vary over an order of magnitude depending on the phase of gas observed. 
\citet{2019ApJ...886..74} (dark blue circles) measure mass-loading factors using \ha\ only and find 
no strong relationship with stellar mass. Rather, they find galaxies with highly concentrated star-formation 
tend to have higher mass-loading factors. \citet{2017MNRAS...469..4831} (light blue crosses) use UV 
absorption lines of extreme starbursts, probing low and high ionization states, and find better agreement 
with simulations. However, these are still consistent with \citet{2019ApJ...886..74} within the uncertainties. 

We find Pox~186 to be markedly different from simulations, and more in line with mass-loading factors 
from \citealt{2019ApJ...886..74} (Figure \ref{fig:massload}). This is understandable as both mass-loading 
factors were derived from only Balmer recombination lines, thus only probing one gas phase. The difference 
between observations and simulations, however, is harder to explain. The results from \citet{2017MNRAS...469..4831} 
suggest observing other phases could start to bring observed mass-loading factors in-line with simulations. 
In addition, measuring mass outflow rates require rough approximations in most cases. It seems unlikely, 
however, that the assumptions could make up for two orders of magnitude difference between Pox~186's mass-loading 
factor and what would be expected from simulations. Likewise, the non-detection
in \hi\ suggests that the mass-loading factor of the ionized gas in Pox~186 plays a more
important role in the total mass loading factor than it may in a normal, more massive galaxy with a large
reservoir of neutral gas \citep{2020ApJ...494..4266}. 

The observed suppressed mass-loading factor can also be the result of a blow-away that significantly depleted the reservoir of gas that can be loaded into the outflow. 
\citet{2002ApJ...581..1039} found the 
central star-cluster to be 4$\pm$1.3 Myrs old using a stellar population synthesis model of single stellar 
populations with various ages assuming instantaneous star-formation. Using Starburst99 models 
\citep{1999ApJS...123..3} and continuous star formation increases this age to 6 Myrs. 
While 4 Myrs is 
young, there is enough time for supernovae to explode and drive significant outflows given that it only takes $\sim10^5$ yrs for the wind, or $\sim10^4$ yrs for 
supernova ejecta to reach the outer edge of Pox~186. 
We can estimate what the mass-loading factor would have been during the blow-away phase as follows:
Assuming the stellar mass to \hi\ mass scaling relation from \citet{2018ApJ...864..40}, Pox~186 would have had 10$^{7.2}$ M$_{\odot}$ in \hi. Including this mass into the outflow over the last 4 Myrs would increase the mass-loading factor to 140, bringing it in line with the extrapolations from simulations at low masses. 
Therefore it is possible that a large 
fraction of the ISM has already been removed by stellar feedback and the suppressed mass-loading factor measured today is a consequence of a previous episode of gas removal. 

In this regard, the simplistic assumption of a single-aged burst of star formation may be misleading us.
\citet{2019Galax...7...88} provides a good overview of the 30~Dor star forming region in the LMC and promotes detailed study of this region as a guide to similar giant star forming complexes which, due to distance, cannot be observed in such detail.  Because the individual stars in the 30~Dor region can be resolved both through imaging and spectroscopy, it is possible to develop an evolutionary history.  \citet{2019Galax...7...88} summarizes that star formation in 30~Dor started 25 Myr ago and has proceeded up until at least 2 Myr ago.  However, if 30~Dor were treated as typical for more distant star forming regions (i.e., as an instantaneous starburst) one would derive a considerably younger age.

\subsection{Pox~186 as a Green Pea Analogue}

In the Introduction, we put forward three observations supporting the view that Pox~186 is a nearby analogue of the Green Pea galaxies.  Our kinematic study has provided a fourth line of evidence.
In a spectroscopic study of six Green Pea galaxies, 
\citet{2012ApJ...754..L22} found high velocity gas (FWZI $\ge$ 1000 km s$^{-1}$) in all of them. Together with the observation by \citet{2017MNRAS...469..4831} that all Green Pea galaxies show evidence of outflows, it seems plausible that outflows are responsible for clearing pathways allowing for the large Lyman continuum escape fractions for the Green Pea galaxies.
In support of this, in an IFU study of Mrk~71, a starburst hosted by the nearby galaxy NGC~2366 which has been deemed the nearest Green Pea analogue by \citet{2017ApJ...845..165}, \citet{2019A&A...623A.145} also find a broad emission line component. \citet{2019A&A...623A.145} conclude that their ``results strongly indicate that kinematical feedback is an important ingredient for LyC leakage in GPs.''  The characteristics of Mrk~71 show many similarities to Pox~186, including a ``wing'' of emission line gas. Thus, in all regards, Pox~186 can be considered as a nearby analogue to the Green Pea galaxies, albeit in a much lower mass regime.

\section{Conclusions} \label{sec:conclusions}

Our new observations of Pox~186 have revealed a remarkable outflow of ionized gas, making a blow-away
scenario increasingly likely. The broad component shows $\sim5-10$\% of the \hii\ mass will be completely 
removed from the galaxy, while the updated distance using Cosmicflows-3 makes Pox~186 even smaller than 
previously thought. It is significant that such a small galaxy ($<1.8 \times$ 10$^7$ \solarmass) is capable 
of producing a star-cluster of 10$^5$ \solarmass, but it remains unclear if the low mass is a result of the 
starburst blowing material out. Assuming this is true we find the mass-loading factor must be $\sim$140 to 
remove the starting reservoir of \hi\ expected from dwarf galaxy scaling relations. Such a large amount of gas 
blown away from the galaxy should still be partly detectable as a halo of gas surrounding Pox~186, much 
like the low surface brightness component present in HST imaging \citep{2002ApJ...581..1039}. Our field 
of view, however, restricts our ability to measure the mass in this component. 

Using MCMC fitting we have investigated the kinematic properties of ionized gas as traced by \oiii\ in
Pox~186. We find the \oiii\ doublet is well fit by a two component model: a strong, narrow, Gaussian
component, and a fainter, broad, Lorentzian component. Across the entire field of view, the narrow component
is present and its flux distribution follows the general light distribution in the galaxy as seen in broad
band imaging, while the broad component is detected in an elongated region extending vertically that is
distinct in shape from the narrow component and continuum. The broad component forms strong evidence for an
outflow in Pox~186 with a velocity 4-14 times larger than the escape velocity. Our spatially resolved 
observations have shown a consistent line-shape across the region, implying some global mechanism is behind 
the formation of the broad component. 

One promising model that simultaneously explains these observations is a scenario where the low density, hot
(T$\sim$10$^6$ K) central starburst outflow emanating from the galaxy flows around high density warm
(T$\sim$10$^4$ K) clouds where it mixes and imparts some of its momentum into the gas on the surface of the
warm clouds. This mixing is what generates the broad wings in the ionized gas while the relatively untouched
core of the clouds emit the narrow component. 

This is not the only viable explanation, however, and follow-up observations are required to validate or
refute this model. It is possible the wing component is due to a series of filaments, clouds, and shells of
ionized gas with a velocity distribution that recreates the wings when projected along the line of sight.
The consistency in line-shape of the broad component would then be an indication of an ISM that is
relatively uniform. The value of this model is that it does not require an unseen phase of gas to be
interacting all the time to recreate the observations. The ionized gas only needs periodic energy and
momentum input from supernova of stars in the central star cluster.  

To distinguish between these two scenarios and confirm or refute the interacting wind model one could look
at a variety of indicators. First, imaging in the FUV and X-ray would immediately show whether a hot unseen
wind phase is present and its radial distribution would indicate whether some of its energy and momentum
is being lost to the warm ionized gas. Second, the turbulent mixing of ionized gas would act to increase
the local magnetic field strength. \citet{2012ApJL...746..L6} find increased magnetic field strengths in GPs 
and suggest their magnetic fields could induce large star-formation events, but in a turbulent mixing scenario 
it could go the other way as well. Synchrotron emission in the radio regime or polarization
observations could be used to measure the magnetic field strength across the galaxy that would then be
compared with the strength of the broad emission. If the wings are indeed the product of mixing then one
would expect increased magnetic field strengths to be correlated with increasing wing strength. 

As a final remark we re-emphasize the unique qualities of Pox~186. The majority of the first galaxies that formed
were likely similar to Pox~186 in mass, being about the size of giant \hii\ regions today, and, like Pox~186 seems to
have done, formed in isolation. Pox~186 offers the closest example of such a galaxy and is a potential
local laboratory to study the extreme ISM properties of the low-mass star-forming galaxies during the early
universe. 
\\

\noindent \textbf{Acknowledgments:} We would like to thank Michele Guala for his insight into turbulent flows, and Kristen McQuinn and John Cannon for providing their data used in this paper. This research made use of NASA/IPAC Extragalactic Database (NED) and NASA's Astrophysical Data System. We also express gratitude to the Gemini Help Desk, which assisted the reduction process. We thank the anonymous referee for their comments which led to significant improvements in our paper.




\appendix \label{appendix}

\begin{figure*}[ht!]
\plotone{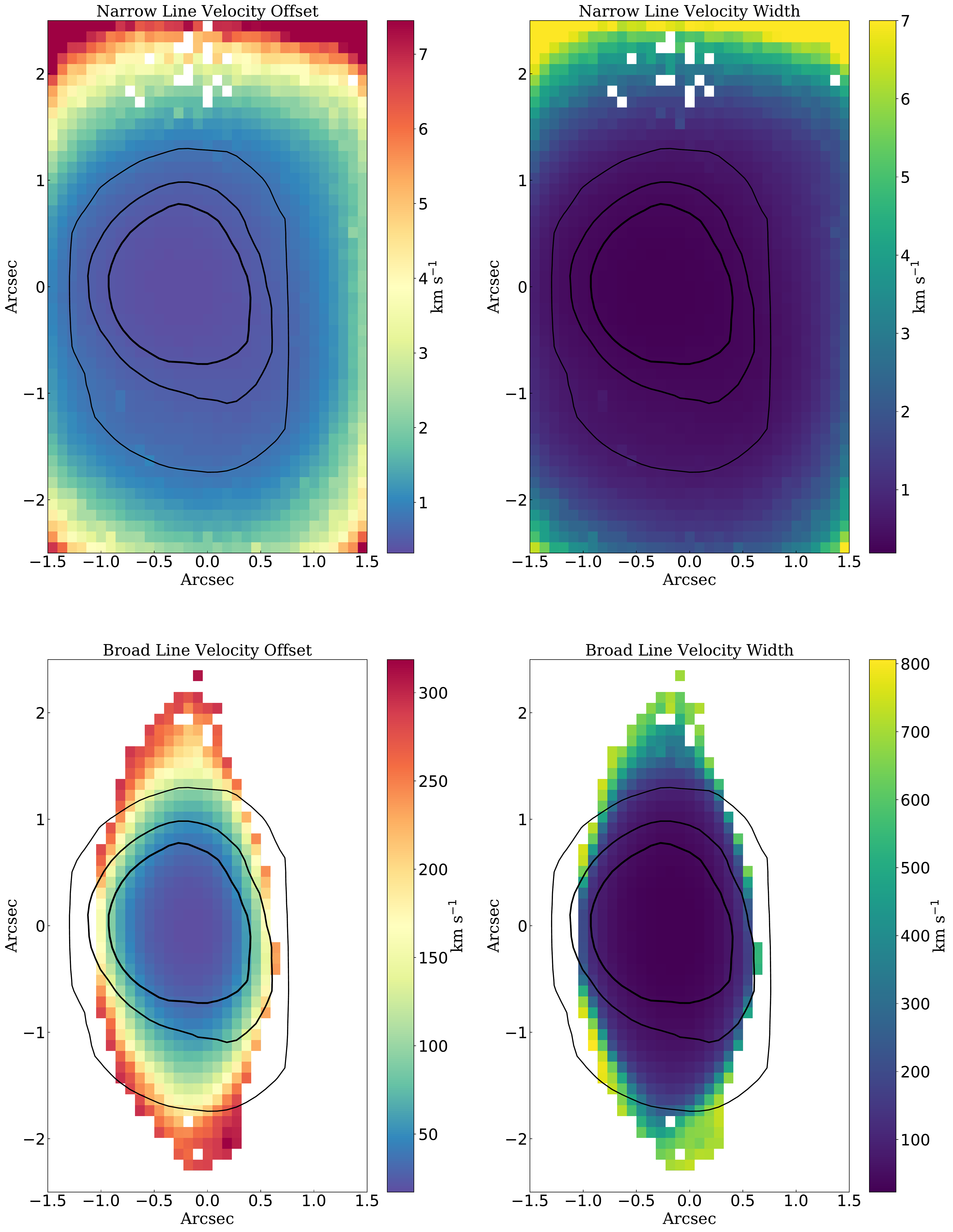} 
\caption{Uncertainties in the kinematics of the two Gaussian components derived from the parameter distribution of two-Gaussian model fit described in Section \ref{sec:data}. The uncertainties shown are the mean of the upper and lower 1$\sigma$ widths of the parameter distributions.\label{fig:mcmcerrs}}
\end{figure*}

\begin{figure*}[ht!]
\plotone{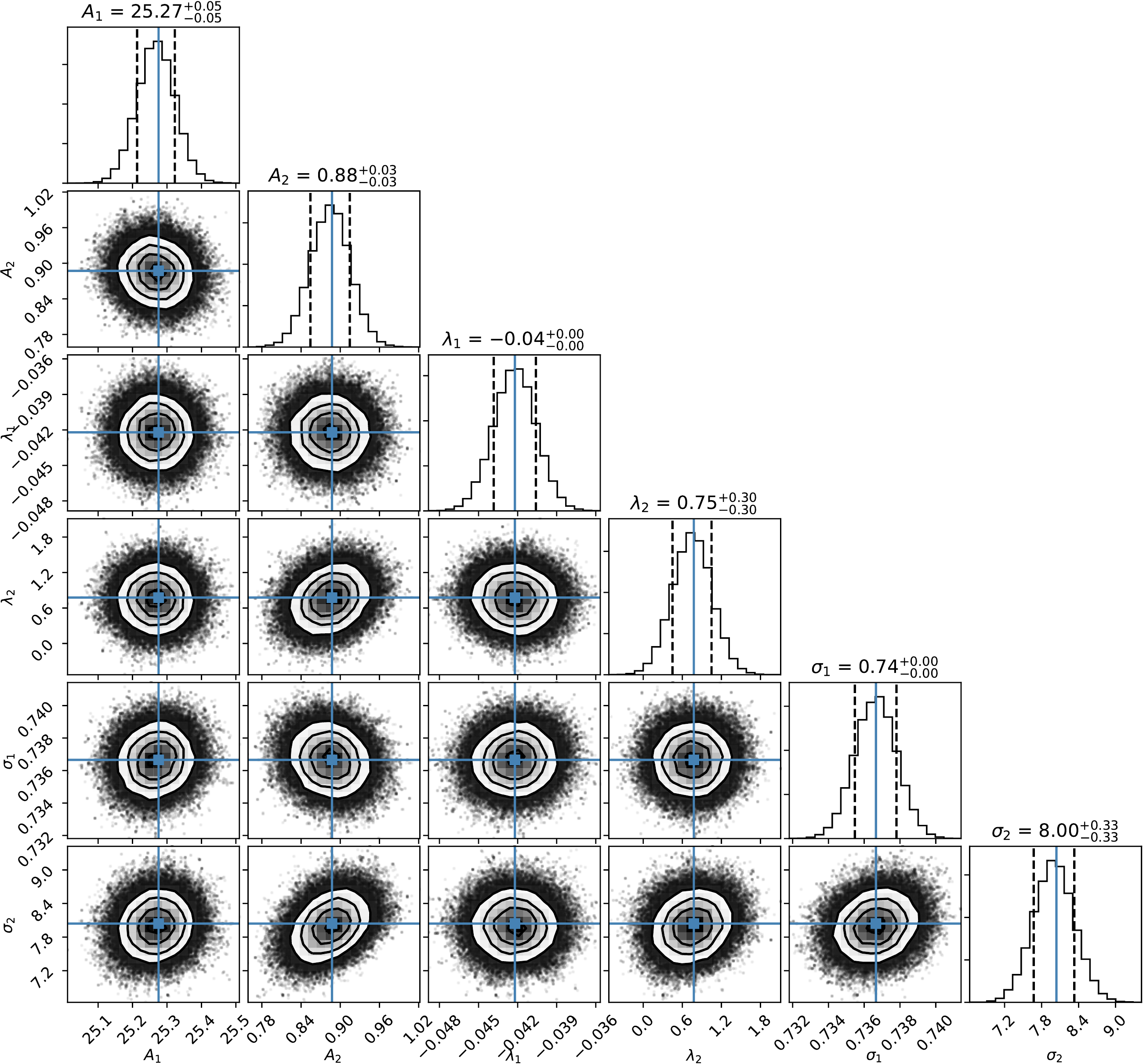}
\caption{One and two dimensional projections of the posterior probability distribution for the Gaussian widths ($\sigma$), line center offsets ($\lambda$), and line fluxes ($A$). The blue lines denote the median of the distributions with the dashed black lines showing the 16\% and 84\% quantiles. The degeneracies between the two components were overcome by restricting the prior range of the amplitude for the broad component. Note: This is a fit to the central, brightest pixel. \label{fig:corner}}
\end{figure*}

\end{document}